\documentclass[12pt]{iopart}

\usepackage{bm}%
\usepackage{graphics}%
\usepackage{graphicx}%
\usepackage{epsfig}%
\usepackage{subfig}
\usepackage{multirow}
\usepackage{float}
\usepackage{dcolumn}
\usepackage{upgreek}
\usepackage{cite}
\usepackage{setspace}
\begin{document}

\title[]{Strong interaction between graphene layer and Fano resonance in terahertz metamaterials}

\author{Shuyuan Xiao$^{1}$, Tao Wang$^{1}$, Xiaoyun Jiang$^{1}$, Xicheng Yan$^{1}$, Le Cheng$^{1}$, Boyun Wang$^{2}$ and Chen Xu$^{3}$}

\address{$^{1}$~Wuhan National Laboratory for Optoelectronics, Huazhong University of Science and Technology, Wuhan 430074, People's Republic of China}
\address{$^{2}$~School of Physics and Electronic-information Engineering, Hubei Engineering University, Xiaogan 432000, People's Republic of China}
\address{$^{3}$~Department of Physics, New Mexico State University, Las Cruces 88001, United State of America}
\ead{wangtao@hust.edu.cn}
\vspace{10pt}
\begin{indented}
\item[]March 2017
\end{indented}

\begin{abstract}
Graphene has emerged as a promising building block in the modern optics and optoelectronics due to its novel optical and electrical properties. In the mid-infrared and terahertz (THz) regime, graphene behaves like metals and supports surface plasmon resonances (SPRs). Moreover, the continuously tunable conductivity of graphene enables active SPRs and gives rise to a range of active applications. However, the interaction between graphene and metal-based resonant metamaterials has not been fully understood. In this work, a simulation investigation on the interaction between the graphene layer and THz resonances supported by the two-gap split ring metamaterials is systematically conducted. The simulation results show that the graphene layer can substantially reduce the Fano resonance and even switch it off, while leave the dipole resonance nearly unaffected, which phenomenon is well explained with the high conductivity of graphene. With the manipulation of graphene conductivity via altering its Fermi energy or layer number, the amplitude of the Fano resonance can be modulated. The tunable Fano resonance here together with the underlying physical mechanism can be strategically important in designing active metal-graphene hybrid metamaterials. In addition, the "sensitivity" to the graphene layer of the Fano resonance is also highly appreciated in the field of ultrasensitive sensing, where the novel physical mechanism can be employed in sensing other graphene-like two-dimensional (2D) materials or biomolecules with the high conductivity.
\end{abstract}

%
%
%
%
%

\section{Introduction}\label{sec1}
Graphene, a monolayer of carbon atoms arranged in plane with a honeycomb lattice, is a two-dimensional (2D) material. Since the first report of its synthesis via a "Scotch tape" method in 2004, graphene has emerged as a popular research topic in the fields of optics and optoelectronics\cite{novoselov2004electric,bonaccorso2010graphene,de2013graphene}. Among its many novel properties, the semi-metallic behavior is one of the most fascinating ones, and based on which, graphene can couple to the incidence light and support surface plasmon resonances (SPRs) in the mid-infrared and terahertz (THz) regime\cite{grigorenko2012graphene,garcia2014graphene,he2014comparison,fan2015tunable,he2016further}. The propagating SPRs in graphene waveguides make it possible to guide light with deep subwavelength mode profiles\cite{nikitin2011edge,lin2015combined,han2015dynamically}, and in the meanwhile, the localized SPRs in graphene metamaterials lead to efficient light confinement and strong near-field enhancement\cite{fan2013enhancing,zhang2015towards,xia2016localized,xiao2016tunable}. Moreover, the continuously tunable conductivity of graphene via manipulating its Fermi energy enables active SPRs\cite{ju2011graphene,he2015tunable,hu2015broadly}, which provide an effective route for efficient real-time control and manipulation of the incidence light and offer much more flexibility than traditional plasmonic materials. Therefore, a range of graphene-based active applications have been proposed and demonstrated in the mid-infrared and THz regime such as absorbers\cite{tahersima2015enhanced,wu2016tunable,linder2016graphene}, biosensors\cite{li2014sensitive,yan2016high}, filters\cite{li2013investigation,li2014tunable,feng2017tunable} and modulators\cite{andersen2010graphene,he2014electrically}, which are considered as potential competitors to their electrically controllable metallic counterparts\cite{chen2006active,fan2017electromagnetic}. Though the function role of graphene in the active control of SPRs has been extensively investigated, the interaction between graphene and metal-based resonant micro/nanostructures has not been fully understood, which is also technically important since it may provide new opportunities to reveal novel physical mechanisms as well as feed back precursors for the calibration of active control and application development of metal-graphene hybrid micro/nanostructures\cite{papasimakis2010graphene,zou2012interaction,dabidian2015electrical,he2015graphene,chakraborty2016gain,fan2016electrically}.

To this end, a simulation investigation on the interaction between the graphene layer and metal-based resonant metamaterials is systematically conducted in this work. The classical THz metamaterials composed of an array of the two-gap split rings is employed here to simultaneously support both the dipole resonance and the Fano resonance. The simulation results show that the presence of the graphene layer on the top of split ring metamaterials can substantially reduce the Fano resonance and even switch it off, while leave the dipole resonance nearly unaffected. The underlying physical mechanism lies in that the conductive graphene layer can recombine and neutralize the opposite type of charges at the ends of the two gaps, and thus suppress the electric field enhancement at the Fano frequency. With the manipulation of graphene conductivity via altering its Fermi energy or layer number, the amplitude of the Fano resonance can be further modulated. Therefore, the tunable Fano resonance here together with the underlying physical mechanism can be strategically important in designing active metal-graphene hybrid metamaterials. In addition, considering the ultra thin thickness of the graphene layer ($\sim 1$ nm), the "sensitivity" of the Fano resonance here is also highly appreciated in the field of ultrasensitive sensing, where the novel physical mechanism can be employed in sensing other graphene-like 2D materials or biomolecules.

\section{The geometric structure and numerical model}\label{sec2}
\begin{figure}[htbp]
\centering
\includegraphics[scale=0.4]{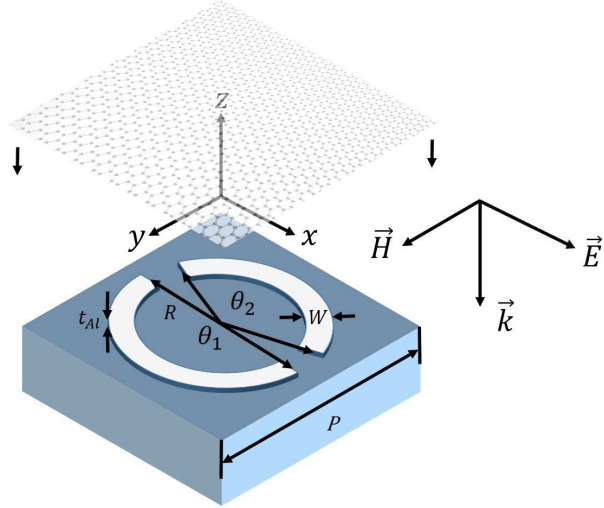}
\caption{\label{fig:1}The schematic geometry of our proposed structure. From the top to the bottom of the structure are a graphene layer, a two-gap aluminum split ring and a silicon substrate. The lattice constant is $P=60$ $\upmu$m. The radius, the width and the thickness of the split ring resonator are respectively $R=21$ $\upmu$m, $W=6$ $\upmu$m and $t_{Al}=200$ nm, and the substrate is assumed to be semi-infinite. The central angles of the lower and the upper arcs of the two-gap split ring are denoted by $\theta_{1}$ and $\theta_{2}$.}
\end{figure}

The schematic geometry of our proposed structure is depicted in Figure~\ref{fig:1}. The unit cell of metal-based resonant metamaterials is arranged in a period array with a lattice constant $P=60$ $\upmu$m and composed of a two-gap aluminum split ring on the top of a silicon substrate. The radius, the width and the thickness of the split ring resonator are respectively $R=21$ $\upmu$m, $W=6$ $\upmu$m and $t_{Al}=200$ nm, and the substrate is assumed to be semi-infinite. The central angles of the lower and the upper arcs of the two-gap split ring are denoted by $\theta_{1}$ and $\theta_{2}$. The optical constants of aluminum in the THz regime are described by a Drude model $\varepsilon_{Al}=\varepsilon_{\infty}-\omega_{p}^{2}/(\omega^{2}+i\omega\gamma)$ with the plasmon frequency $\omega_{p}=2.24\times 10^{16}$ rad/s and the damping constant $\gamma=1.22\times 10^{14}$ rad/s\cite{ordal1985optical}. The refractive index of the silicon is taken as $n_{Si}=3.42$.

The graphene layer is placed on the top of the two-gap split ring metamaterials and modeled as a 2D flat plane. The graphene conductivity in the THz regime is dominated by the intraband transport processes and also treated according to a Drude-like model $\sigma_{g}=i e^{2}E_{F}/[\pi\hbar^{2}(\omega+i/\tau)]$, where $e$ is the charge of an electron, $E_{F}$ is the Fermi energy of graphene, $\hbar$ is the reduced Planck's constant and $\tau$ is the relaxation time. $E_{F}=0.275$ eV and $\tau=70$ fs can be calculated by $E_{F}=\hbar v_{F}\sqrt{\pi|n_{g}|}$ and $\tau=\mu\hbar\sqrt{\pi|n_{g}|}/e v_{F}$, where the Fermi velocity $v_F=1.1\times 10^{6}$ m/s, the charge density $n_{g}=4.60\times 10^{12}$ cm$^{-2}$ and the carrier mobility $\mu=3000$ cm$^{2}$/V$\cdot$s are consistent with the experimental measurements.
\cite{zhang2005experimental,jnawali2013observation}.

As previously reported, the two-gap split ring metamaterials can simultaneously support both a bright dipole resonance and a dark Fano resonance if the symmetry of the structure is broken\cite{singh2011sharp,singh2013fano}, which can provide an excellent platform for the comparative investigation on the interaction between the graphene layer and THz resonances. In the initial setup, the central angles of the lower and the upper arcs of the two-gap split ring are equally set to $\theta_{1}=\theta_{2}=160^{\circ}$ to form symmetric split ring (SRR) metamaterials. The asymmetry is gently introduced with the increase of $\theta_{1}$ and the decrease of $\theta_{2}$ to make asymmetric split ring (ASR) metamaterials, and the asymmetry degree is defined as $\Delta\theta=\theta_{1}-\theta_{2}$. The influence of the graphene layer on the THz resonances with different $\Delta\theta$ will be analyzed to reveal the novel physical mechanism behind it using the finite-difference time-domain (FDTD) method. The periodical boundary conditions are employed in the $x$ and $y$ directions and perfectly matched layers are utilized in the $z$ direction along the propagation of the incidence plane wave.

\section{Simulation results and discussions}\label{sec3}
\begin{figure}[htbp]
\centering
\includegraphics[scale=0.8]{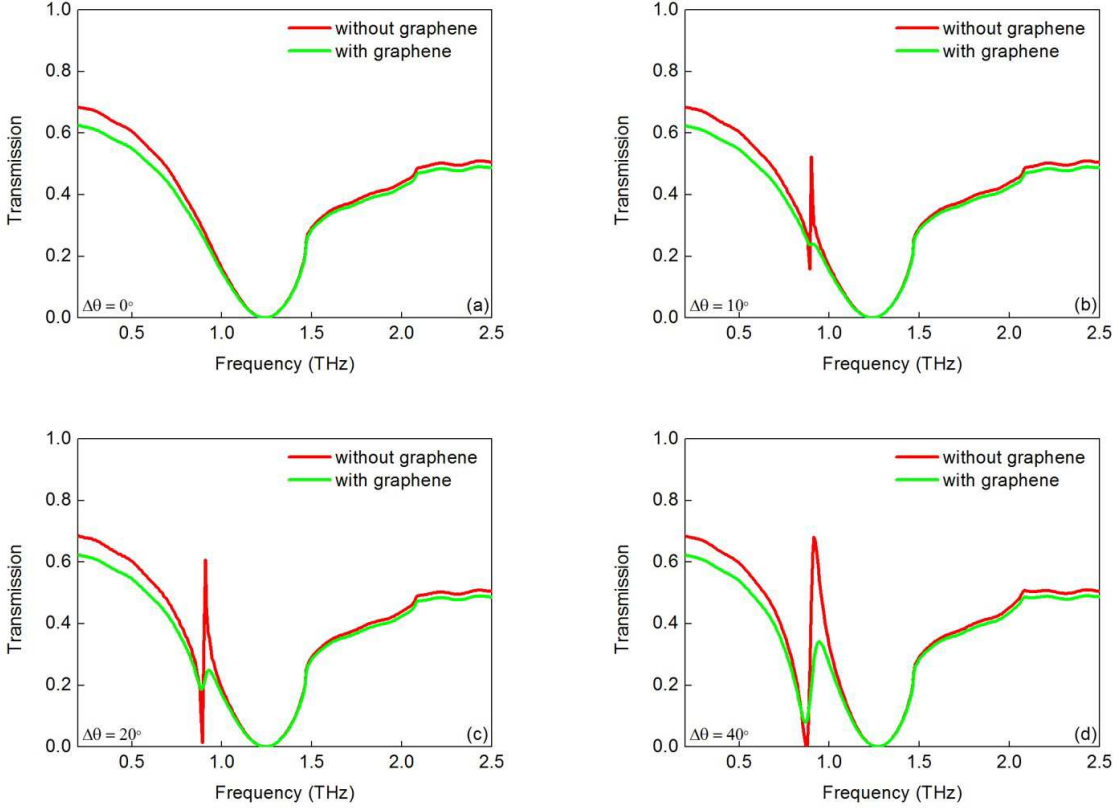}
\caption{\label{fig:2}The simulated transmission spectra through the two-gap split ring metamaterials without and with the graphene layer. The THz plane wave is illuminated along the negative $z$-axis with the electric field oriented perpendicular to the gaps. The asymmetry degree is varied from 0$^{\circ}$ to 40$^{\circ}$, as shown in the insets.}
\end{figure}

The THz plane wave is illuminated along the negative $z$-axis with the electric field oriented perpendicular to the gaps. The simulated transmission spectra through the two-gap split ring metamaterials without and with the graphene layer is shown in Figure~\ref{fig:2}, where the asymmetry degree is varied from 0$^{\circ}$ to 40$^{\circ}$.

In the absence of the graphene layer, with the asymmetry degree $\Delta\theta=0^{\circ}$, the SSR metamaterials show a symmetric resonance at 1.24 THz, which is known as the bright dipole mode. It is the basic and direct response of metamaterials to the free-space light and exhibits a broad resonance line width caused by the strong radiative losses. When a slight asymmetry $\Delta\theta=10^{\circ}$ is introduced, SSR metamaterials turn into ASR ones, along with two resonances appearing in the transmission spectrum, one which is identical to that of SSR metamaterials and recognized as the dipole mode, the other is an asymmetric Fano resonance located at 0.89 THz. This additional ultra sharp resonance results from a subradiative dark mode where the radiative losses are near completely suppressed and the line width of the transmission profile is solely limited by the intrinsic metal losses (Drude damping)\cite{liu2009planar}. With the increase of the asymmetry degree, the dipole resonance has little change except the resonance frequency shifting to 1.27 THz, which once again demonstrate the fundamentality of this bright mode in the symmetric as well as the asymmetric metamaterials, and in the meanwhile, the Fano resonance remains at the nearly fixed frequency and becomes even more pronounced, where the steeper dips can be nicely observed in the respective spectra with $\Delta\theta=20^{\circ}$ and $\Delta\theta=40^{\circ}$.

In the presence of the graphene layer, the profile of the broad dipole resonance is pretty much the same as that in the previous case with every asymmetry degree, which indicates that the bright mode does not interact with the graphene layer. By contrast, however, the Fano resonance experiences a great change. Though the trend of the increase in the amplitude of the Fano resonance is retained with the increase of asymmetry degree, the absolute strength is substantially reduced and even switched off. In order to characterize the induced change in the resonance strength due to the presence of the graphene layer, the reduction degree in the transmission is introduced as $\Delta T=|T_{0}-T_{g}|\times 100\%$, where $T_{0}$ and $T_{g}$ are the transmission amplitudes at the dip of the Fano resonance without and with the graphene layer. When the asymmetry degree starts at $\Delta\theta=10^{\circ}$, the reduction degree in the transmission is $\Delta T_{10}=7.93\%$. As $\Delta\theta$ increases to 20$^{\circ}$, the reduction degree goes up to $\Delta T_{20}=17.28\%$. Finally when $\Delta\theta$ comes to 40$^{\circ}$, the reduction degree gets back to $\Delta T_{40}=7.33\%$. In comparison with that on the dipole resonance, the graphene layer has a much more pronounced effect on the Fano resonance, which implies the existence of a strong interaction between the graphene layer and ASR metamaterials at the Fano frequency.

\begin{figure}[htbp]
\centering
\includegraphics[scale=0.8]{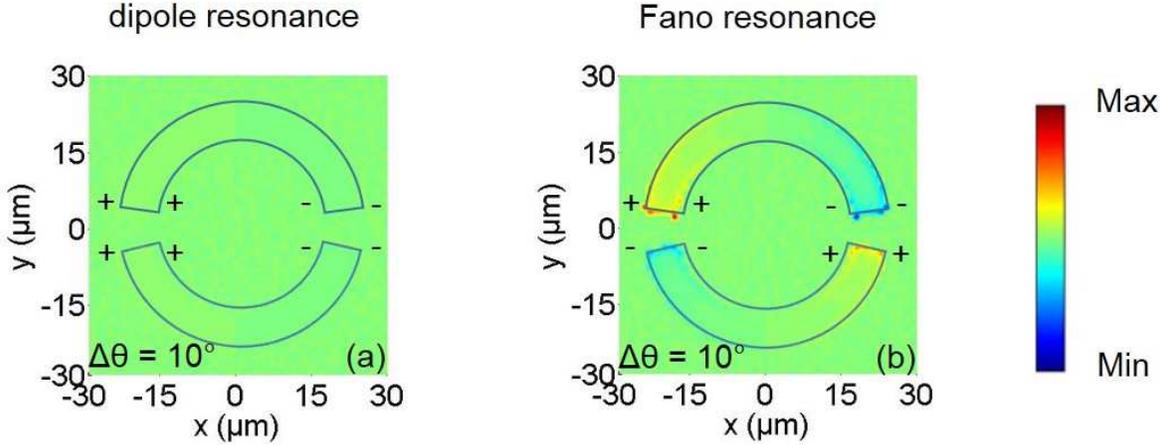}
\caption{\label{fig:3}The simulated surface charge density distributions (a) at the dipole resonance and (b) at the Fano resonance without the graphene layer. The asymmetry degree is $10^{\circ}$, as shown in the insets.}
\end{figure}

\begin{figure}[htbp]
\centering
\includegraphics[scale=0.8]{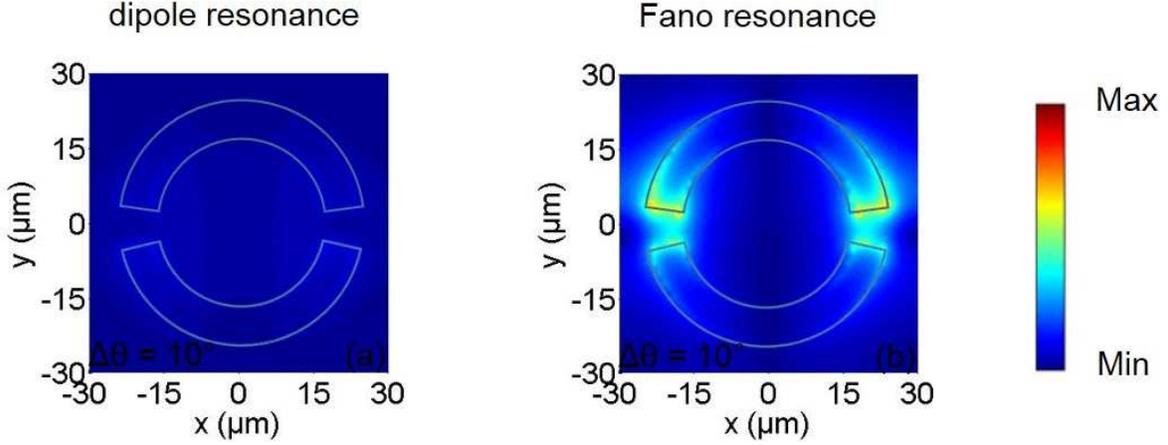}
\caption{\label{fig:4}The simulated electric field distributions (a) at the dipole resonance and (b) at the Fano resonance without the graphene layer. The asymmetry degree is $10^{\circ}$, as shown in the insets.}
\end{figure}

To reveal the physical mechanism behind the novel phenomenon, the essences of the dipole and the Fano resonances are reviewed. The simulated surface charge density and electric field distributions at the resonances without the graphene layer are plotted in Figure~\ref{fig:3} and \ref{fig:4}, where the asymmetry degree is $\Delta\theta=10^{\circ}$. For the dipole resonance at 1.24 THz, a vertically symmetric distribution of the surface charge density along the lower and upper arcs of ASR can be nicely observed and the net induced dipoles in the THz metamaterials exhibit a clear tendency to follow the polarization of the incidence plane wave, where the same type of charges locate at two ends of each gap. However, for the Fano resonance at 0.89 THz, a large amount of the opposite type of charges accumulate at the two gaps, inferring a circulation distribution of the surface charge density along the entire ASR and making it a magnetic dipole, which can be phenomenologically understood in terms of a $LC$-oscillator with the two gaps serving as effective capacitors. Therefore, much more pronounced electric field are confined in the split gaps of ASR at the Fano resonance since the asymmetric mode is only weakly coupled to the free-space light. By comparison, the different distribution types should be the reason why the two resonances behave differently to the graphene layer. The underlying physical mechanism lies in the high conductivity of graphene. Once the graphene layer is placed on the top of the THz metamaterials, it connects the lower and upper arcs of ASR, shorten the two gaps and thus influences the charge distributions. For the dipole resonance, the presence of the graphene layer does not cause a significant change in the transmission spectrum in that the shorting of the two gaps would not influence the accumulation of the same type of charges. For the Fano resonance, however, the opposite type of charges at the ends of the two gaps can be recombined and neutralized through the highly conductive graphene, which leads a strong suppression of the electric field enhancement. Therefore, the absolute strength of the Fano resonance would be substantially reduced and even switched off in the transmission spectrum.

\begin{figure}[htbp]
\centering
\includegraphics[scale=0.8]{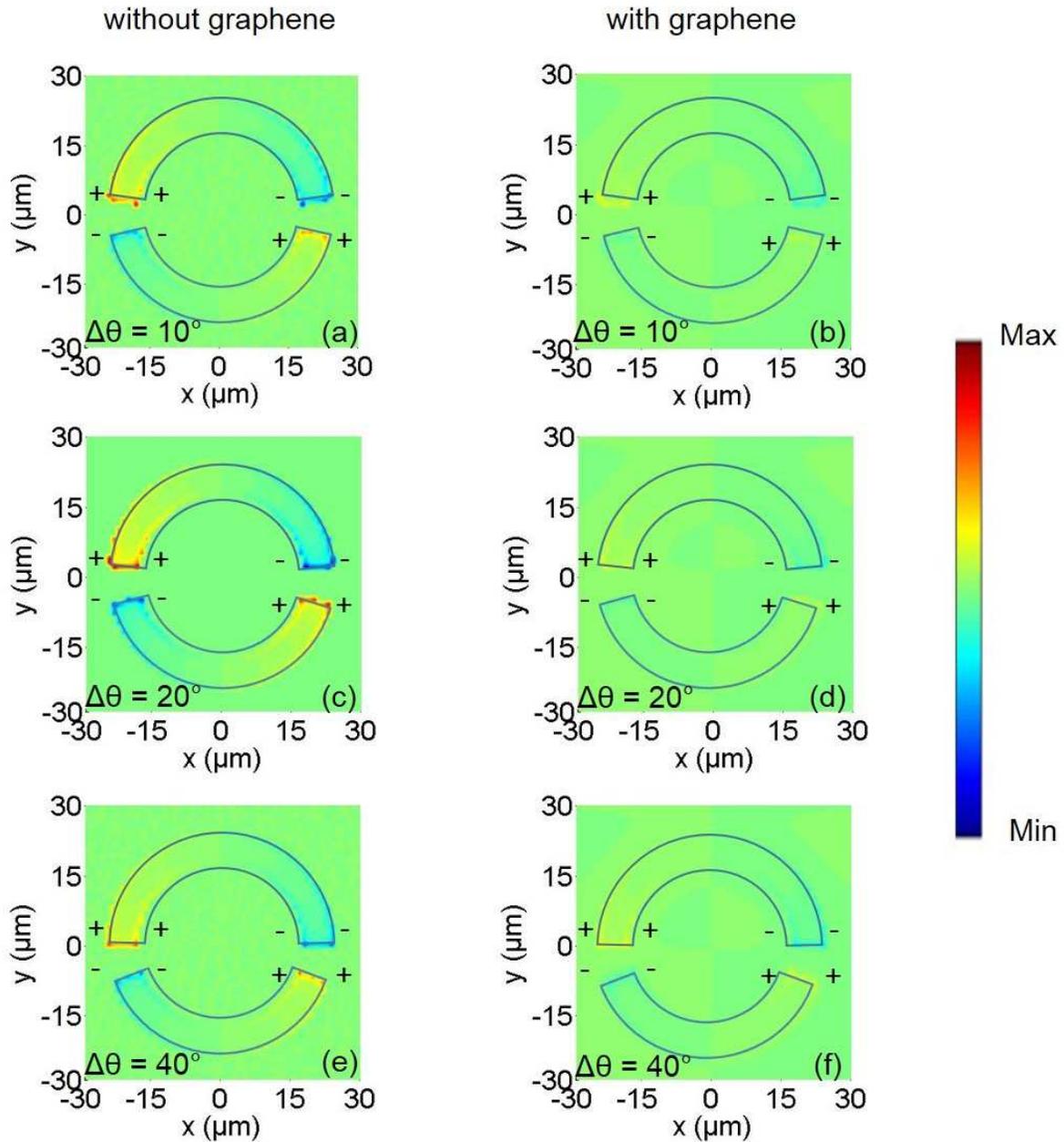}
\caption{\label{fig:5} The simulated surface charge density distributions at the Fano resonance (a-e) without and (b-f) with the graphene layer. The asymmetry degree is varied from $10^{\circ}$ to 40$^{\circ}$, as shown in the insets.}
\end{figure}

\begin{figure}[htbp]
\centering
\includegraphics[scale=0.8]{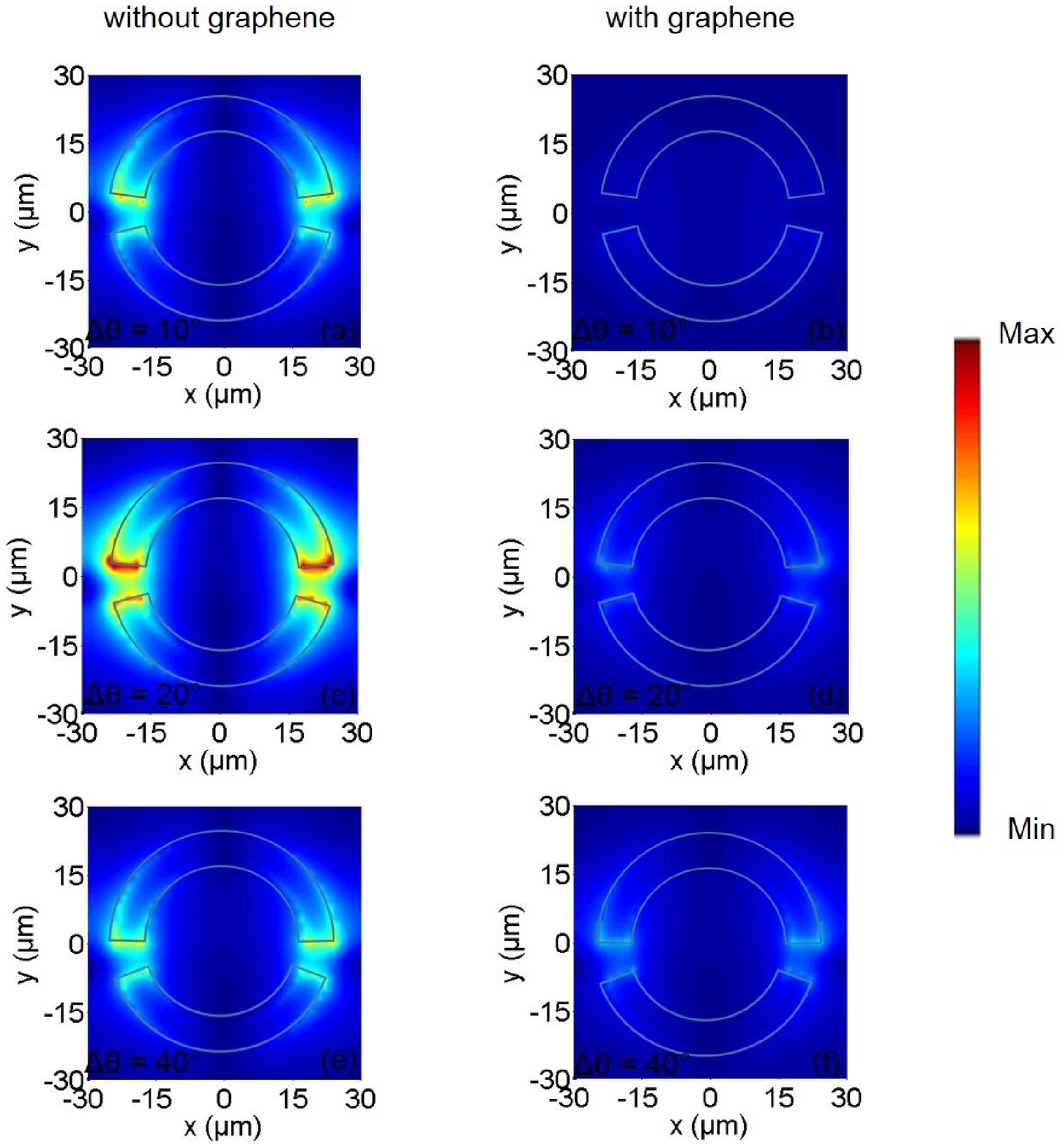}
\caption{\label{fig:6} The simulated electric field distributions at the Fano resonance (a-e) without and (b-f) with the graphene layer. The asymmetry degree is varied from $10^{\circ}$ to 40$^{\circ}$, as shown in the insets.}
\end{figure}

To demonstrate the proposed physical mechanism, the simulated surface charge density and electric field distributions at the Fano resonance without and with the graphene layer are compared in Figure~\ref{fig:5} and \ref{fig:6}, where the asymmetry degree is varied from 10$^{\circ}$ to 40$^{\circ}$. When the asymmetry degree starts at $\Delta\theta=10^{\circ}$, the opposite type of charges at the ends of the two gaps are recombined and neutralized through the conductive graphene placed on the top of ASR metamaterials and the strong enhancement of the electric field in the gaps is near completely suppressed, which corresponds to the switch-off of the Fano resonance in the transmission spectrum. With the increase of asymmetry degree to $\Delta\theta=20^{\circ}$, the surface charge distribution gets denser and the electric field enhancement becomes even more pronounced without the graphene layer, while those with the presence of the graphene layer increase only slightly, leading to the larger change in the absolute strength of the Fano resonance. Finally when $\Delta\theta=40^{\circ}$, the surface charge density and the electric field enhancement begin to fall, and in the meanwhile, those with graphene further go up, which accordingly explains the decrease of the reduction degree in the transmission.

\begin{figure}[htbp]
\centering
\includegraphics[scale=0.8]{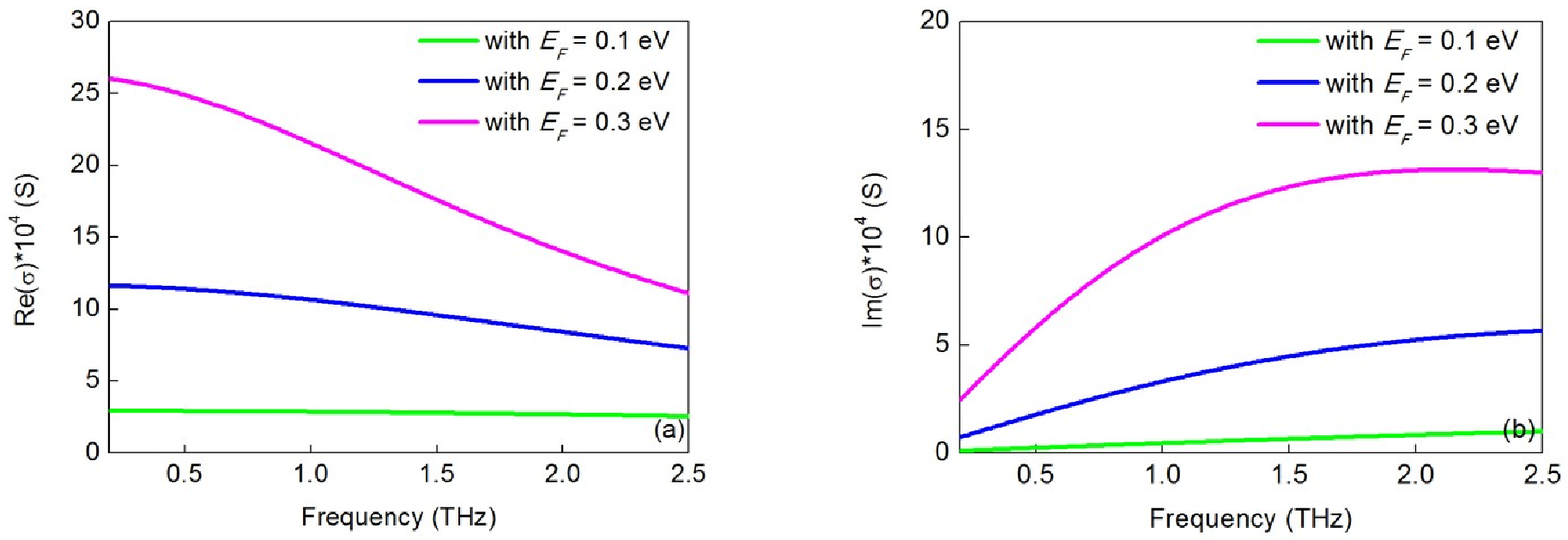}
\caption{\label{fig:7} The frequency dependent graphene conductivity (a) the real part (b) the imaginary part. The Fermi energy is varied from 0.1 eV to 0.3 eV, as shown in the insets.}
\end{figure}

\begin{figure}[htbp]
\centering
\includegraphics[scale=0.8]{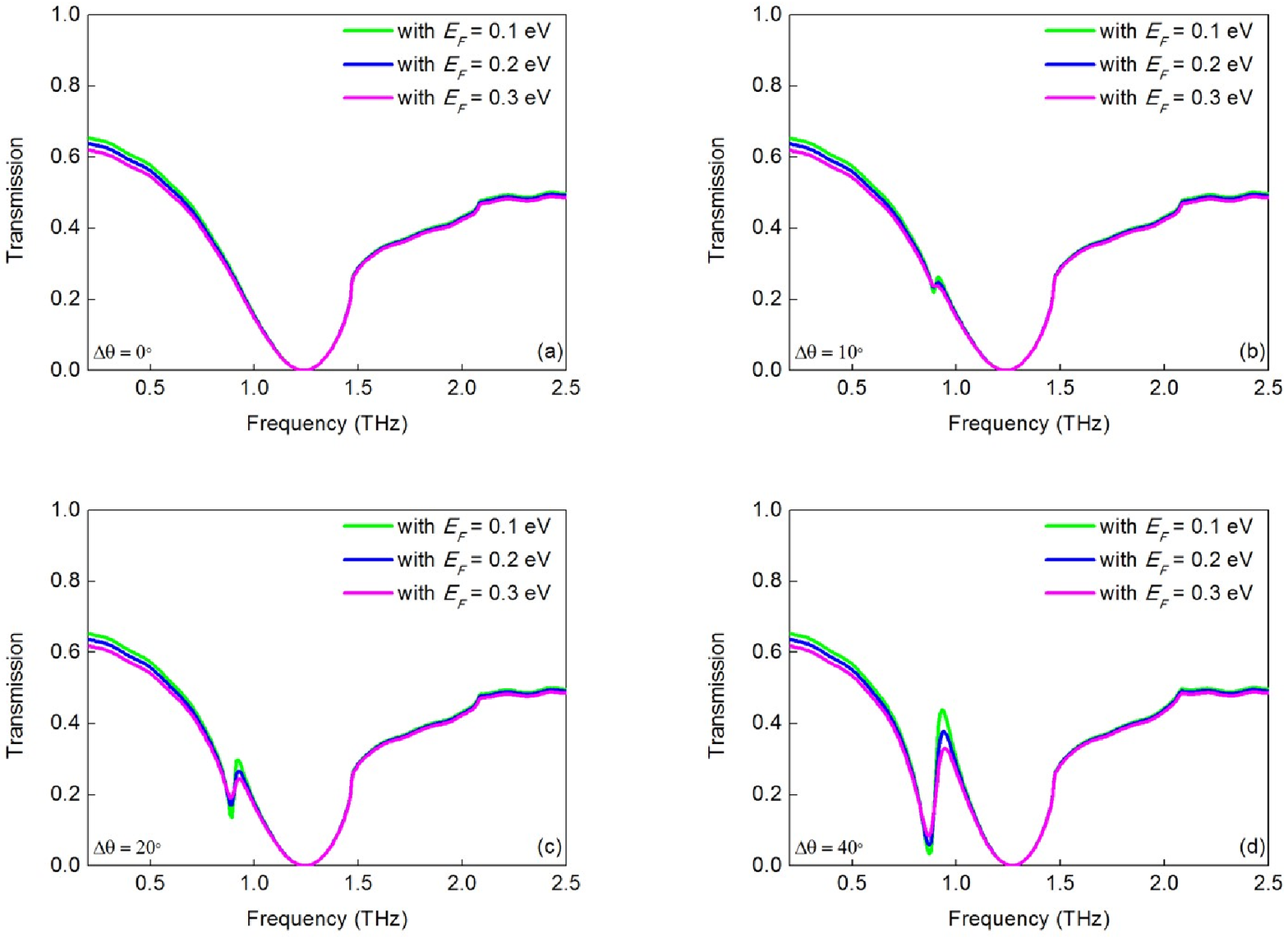}
\caption{\label{fig:8} The simulated transmission spectra through the two-gap split ring metamaterials with the graphene layer. The Fermi energy is varied from 0.1 eV to 0.3 eV and the asymmetry degree from 0$^{\circ}$ to 40$^{\circ}$, as shown in the insets.}
\end{figure}

Compared with metal-based micro/nanostructures, one of the main advantages of metal-graphene hybrid metamaterials is the active tunability. In previous reports, the manipulation of graphene conductivity has been demonstrated via altering the Fermi energy by electric gating\cite{hu2015broadly} or chemical doping\cite{liu2011chemical}. For this purpose, the influence of graphene conductivity on the Fano resonance is also investigated here via altering the Fermi energy of the graphene layer. The frequency dependent graphene conductivity at relatively low Fermi energy is calculated according to the Drude-like model and shown in Figure~\ref{fig:7}. With the increasing Fermi energy from $E_{F}=0.1$ eV to $E_{F}=0.3$ eV, the graphene conductivity increases rapidly, which infers that the recombination effect of the opposite charges at the two ends of the split can be more pronounced with higher Fermi energy.
Considering this, the transmission spectra through the two-gap split ring metamaterials with the graphene layer are simulated in Figure~\ref{fig:8}, where the Fermi energy is varied from 0.1 eV to 0.3 eV and the asymmetry degree from 0$^{\circ}$ to 40$^{\circ}$.
With the increase of the Fermi energy, the amplitude of the Fano resonance with every asymmetry degree decreases, which exhibits the same tendency in accordance with the relationship between conductivity and the Fermi energy of graphene and in turn demonstrate our proposed physical mechanism. Therefore the tunable Fano resonance here together with the underlying physical mechanism can be of strategic importance and practical significance to feed back precursors for the calibration of active control and application development of metal-graphene hybrid metamaterials.

\begin{figure}[htbp]
\centering
\includegraphics[scale=0.8]{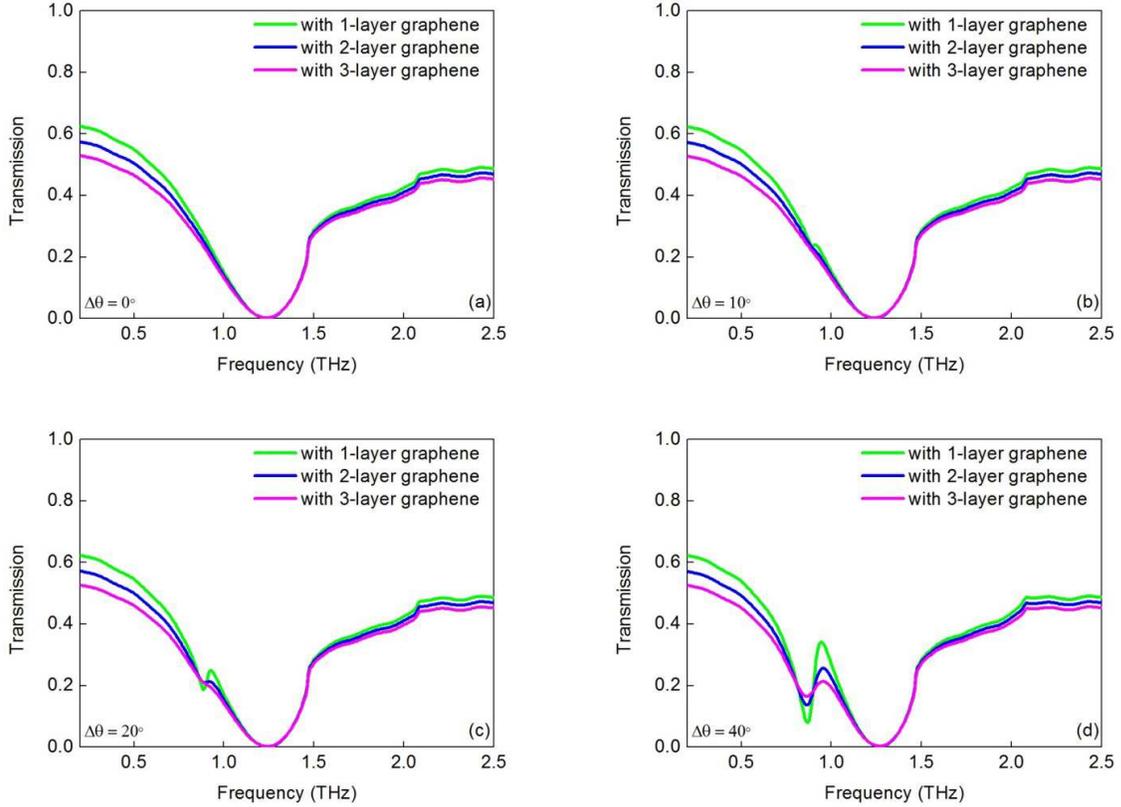}
\caption{\label{fig:9} The simulated transmission spectra through the two-gap split ring metamaterials with the multilayer graphene. The layer number is varied from 1 to 3 and the asymmetry degree from 0$^{\circ}$ to 40$^{\circ}$, as shown in the insets.}
\end{figure}

In the real cases, there would be unavoidable disorders generated by the environment during the growth or transfer processes of graphene. The multilayer graphene can be found either in twisted configurations where the layers are rotated relative to each other or graphitic Bernal stacked configurations where half the atoms in one layer lie on half the atoms in others\cite{brown2012twinning}. Previous investigations have claimed the randomly stacked graphene layers would still behave as the isolated graphene layer owing to the electrical decoupling and the conductivity becomes proportional to the layer number\cite{hass2008multilayer,li2016monolayer}. Considering this, the transmission spectra through the two-gap split ring metamaterials with the multilayer graphene are also simulated in Figure~\ref{fig:9}, where the layer number is varied from 1 to 3 and the asymmetry degree from 0$^{\circ}$ to 40$^{\circ}$.
As with the graphene layer, the dipole resonance still does not interact with the multilayer graphene in that the shorting of the two gaps would not influence the accumulation of the same type of charges. But on the other hand, the presence of the multilayer graphene further reduces the absolute strength of the Fano resonance in the transmission spectra due to the higher conductivity, where this resonance mode is near completely switched off with $\Delta\theta=10^{\circ}$ and $\Delta\theta=20^{\circ}$ and the steepest dip with $\Delta\theta=40^{\circ}$ become quite moderate. These simulation results are in accordance with the relationship between conductivity and the layer number of graphene and once again demonstrate our proposed physical mechanism. In addition, the "sensitivity" to the graphene layer of the Fano resonance in the ASR metamaterials can also be employed in the field of ultrasensitive sensing. In general, sensing with THz metamaterials requires the analyte with a thickness of hundreds of nanometers\cite{o2008thin,tao2010performance}, however, in our case, the ultra thin ($\sim 1$ nm) graphene layer is detected, which can be extended to other graphene-like 2D materials or biomolecules with the high conductivity.

\section{Conclusions}\label{sec4}
In conclusions, the interaction between the graphene layer and THz resonances is numerically investigated in this work. It is found that the presence of the graphene on the top of ASR metamaterials can substantially reduce the Fano resonance and even switch it off, while leave the dipole resonance nearly unaffected. This novel phenomenon is well explained with the high conductivity of graphene, which can recombine and neutralize the opposite type of charges at the ends of the two gaps, and leads a strong suppression of the absolute strength of the Fano resonance. With the manipulation of graphene conductivity via altering its Fermi energy, the amplitude of the Fano resonance can be actively modulated. The multilayer graphene as disorders in the real cases are also included to explore the practical possibility and the results are in accordance with the relationship between conductivity and the layer number of graphene. Therefore the tunable Fano resonance here together with the underlying physical mechanism can be strategically important in designing active metal-graphene hybrid metamaterials. In addition, the "sensitivity" to the graphene layer of the Fano resonance in the ASR metamaterials is also highly appreciated in the field of ultrasensitive sensing, where the novel physical mechanism can be employed in sensing other graphene-like 2D materials or biomolecules with the high conductivity.

\section*{Acknowledgments}
The author Shuyuan Xiao (SYXIAO) expresses his deepest gratitude to his Ph.D. advisor Tao Wang for providing guidance during this project. SYXIAO would also like to thank Dr. Qi Lin, Dr. Guidong Liu and Dr. Shengxuan Xia (Hunan University) for beneficial discussions on graphene optical properties. This work is supported by the National Natural Science Foundation of China (Grant No. 61376055, 61006045 and 11647122), the Fundamental Research Funds for the Central Universities (HUST: 2016YXMS024) and the Project of Hubei Provincial Department of Education (Grant No. B2016178).

\end{document}